\documentclass[10pt,letterpaper,twocolumn]{article} %% two column, final layout

\usepackage{ol2}
\usepackage[draft]{hyperref}
\usepackage{amsmath}
\usepackage{graphicx}
\usepackage{amsmath}
\usepackage{amssymb}
\usepackage{url}
\usepackage{bm}
\usepackage[font={small}]{caption}

\begin{document}

\twocolumn[ %% activate for two-column option

\title{Effect of hole-shape irregularities on photonic crystal waveguides}

\author{Momchil Minkov$^{1,*}$ and Vincenzo Savona$^1$}

\address{$^1$Institute of Theoretical Physics, Ecole Polytechnique Fédérale de Lausanne EPFL, CH-1015 Lausanne, Switzerland
$^*$Corresponding author: momchil.minkov@epfl.ch}

\begin{abstract}

The effect of irregular hole shape on the spectrum and radiation losses of a photonic crystal waveguide is studied using Bloch-mode expansion. Deviations from a perfectly circular hole are characterized by a radius fluctuation amplitude and correlation angle. It is found that the parameter which determines the magnitude of the effect of disorder is the standard deviation of the hole areas. Hence, for a fixed amplitude of the radius fluctuation around the hole, those effects are strongly dependent on the correlation angle of the irregular shape. This result suggests routes to potentially improve the quality of photonic crystal structures. 
%The effect of irregular hole shape on the spectrum and radiation losses of a photonic crystal waveguide is studied using Bloch-mode expansion. Deviations from a perfectly circular hole are characterized by a radius fluctuation amplitude and correlation angle. It is found that the parameter which determines the magnitude of the effect of disorder is the standard deviation of the hole areas. Hence, for a fixed amplitude of the radius fluctuation, those effects are strongly dependent on the correlation angle of the hole shape. This result suggests routes to potentially improve the quality of photonic crystal structures. 
\end{abstract}

 ] %% activate for two-column option

Photonic crystals (PHCs) lie at the forefront of research in photonics \cite{Joannopoulos_2008}. One of the main obstacles to the practical implementation of PHCs is disorder originating from the fabrication process. Disorder sets a severe limitation to the quality factor of high-Q PHC cavities \cite{Gerace_PAN_2005, Noda_OE_2011, Portalupi_PRB_2011}, and degrades the preformance of PHC waveguides by inducing extrinsic radiation losses \cite{Gerace_OL_2004,Hughes_PRL_2005,Hughes_PRB_2009,Mazoyer_PRL_2009} and light localization \cite{John_PRL_1987} in the vicinity of the guided band edge. For this reason, disorder in PHCs has been the subject of increasingly intense theoretical \cite{Gerace_PAN_2005, Gerace_OL_2004,Hughes_PRL_2005,Hughes_PRB_2009,Mazoyer_PRL_2009, Savona_PRB_2011, Portalupi_PRB_2011, Noda_OE_2011, Hughes_PRB_2010} and experimental \cite{Portalupi_PRB_2011, Noda_OE_2011, houdre, Mazoyer_OE_2010, Le_Thomas_PRB_2009} studies in the last decade. Several theoretical works \cite{Gerace_OL_2004, Mazoyer_PRL_2009, Savona_PRB_2011, Portalupi_PRB_2011,Noda_OE_2011} 
have considered the simplest possible disorder model, namely circular holes with randomly fluctuating radii and/or positions, and studied radiation losses, spectral properties, and light localization as a function of the fluctuation amplitude. An analysis based on more realistic assumptions, clarifying the role of the observed deviations from a perfectly circular shape of the PHC holes \cite{skorobogatiy, houdre}, remains a major challenge. Recent experiments \cite{houdre} have provided strong evidence that the relevant parameter, that quantifies the effect of disorder, is the amplitude of {\em fluctuations in the hole area}. On the theoretical side significant progress has been made \cite{Gerace_PAN_2005, Hughes_PRL_2005, Hughes_PRB_2009, Hughes_PRB_2010}, in particular by providing an estimate of the disorder-induced frequency shift of the band edge \cite{Hughes_PRB_2010}, and by showing that the correlation angle of the hole shape -- the measure of how rapidly the radius fluctuates around a hole -- strongly affects the loss rates in a PHC waveguide for a fixed amplitude of the radius fluctuations \cite{Hughes_PRL_2005}. Yet, the interplay of angle correlation and radius fluctuations, and the relevance of hole-area fluctuations, are still poorly understood.

In this letter we address this problem by employing the recently developed Bloch-mode expansion (BME) method \cite{Savona_PRB_2011}. BME is a non-perturbative mode expansion that converges to the exact solution of Maxwell equations within the desired spectral range, while requiring a moderate computational effort, and that allows to implement the hole-shape model of choice in a semi-analytical fashion. As a prototypical system we study a W1-defect waveguide, but our conclusions about the effect of irregular hole shape have a much more general extent. We carry out a systematic analysis for two limiting assumptions within the hole-shape model, namely fixed radius fluctuation (within each single hole), or fixed hole area fluctuation (among different holes). The result clearly shows that varying the correlation angle has practically no effect on spectrum and radiation losses in the second case, while it is highly determinant within the first assumption. We conclude by clarifying the mechanism underlying this phenomenology and discuss how it relates to the fabrication process. 

We consider a triangular lattice of cylindrical holes of radius $R = 0.3a$ in a dielectric slab of thickness $d = 0.5a$, where $a$ is the lattice parameter. The structure presents a ``missing'' row of holes (i.e. a W1 waveguide). The permittivity is set to $\varepsilon_2 = 12.11$ in the dielectric material and to $\varepsilon_1 = 1$ outside. Disorder is modeled in the form of fluctuations in the hole profile, given by $R(\phi) = R + \delta R(\phi)$, where $\phi$ is the polar angle relative to the hole center. In particular, each hole is characterized by a different disorder realization (no correlation between hole shapes is assumed) defined by random Fourier expansion coefficients $C_m$ as 

\begin{equation}
 \delta R(\phi) = \sum_{m=-\infty}^{+\infty} C_m e^{im\phi}.
\end{equation}

The Bloch modes of the regular W1 waveguide are computed with the guided-mode expansion method \cite{andreani_2006}. In all computations we truncate the reciprocal space of guided modes to $G_{max} = 3(2\pi /a)$. To apply BME, we need the Fourier transforms of the dielectric profiles of both the regular structure, $\varepsilon(\mathbf{r})$, and the disordered one, $\varepsilon'(\mathbf{r})$. We take the ${\bf z}$-axis to be orthogonal to the slab plane, in which case both profiles are piecewise constant along ${\bf z}$. The only nontrivial task is then the evaluation of the 2-D Fourier transforms of $\varepsilon(\bm{\rho})$ and $\varepsilon'(\bm{\rho})$ at $z=0$ inside the slab. The former is readily computed \cite{Savona_PRB_2011}, while the latter we compute analytically to first order in $\delta R$. We denote the in-plane position of the center of each hole by $\bm{\rho}_{\zeta}$, with $\zeta$ running over all holes in the structure, the area of the structure by $S$, $g = |\mathbf{g}|$, and the $m$-th Bessel function of the first kind by $J_m(x)$. Then, 

\begin{align}
 &\varepsilon'(\mathbf{g}) =  \varepsilon(\mathbf{g}) + \label{epsprime} \\ \nonumber
 &\frac{2 \pi R_0 (\varepsilon_1 - \varepsilon_2)}{A} \sum_{\zeta}\mathrm{e}^{i\mathbf{g}\bm{\rho}_{\zeta}}\sum_{m = -\infty}^{\infty} i^m \mathrm{e}^{im\theta} C_{m, \zeta} J_m(gR_0).
\end{align}

Here, we always take the $\{C_{m}\}$ coefficients for each hole to be Gaussian random variables, whose distribution is fully determined by $\langle C_m \rangle$ and $\langle |C_m|^2 \rangle$. Note that $\langle C_{m \neq 0} \rangle = 0$, since it is an average over complex numbers with a random phase. We further assume Gaussian correlation along $\phi$, such that $\langle \delta R(\phi) \delta R(\phi')\rangle = \sigma^2 e^{-\frac{(\phi - \phi')^2}{2 \delta^2}}$, where $\delta$ is the correlation angle. This sets $\langle \sum_m |C_m|^2 \rangle = \langle \delta R^2 \rangle = \sigma^2$, and the dependence with $m$ of the second moments to

\begin{equation}
 \langle |C_m|^2 \rangle = \sigma^2 \int_{-\pi}^{\pi} e^{-\frac{\phi^2}{2 \delta^2}} e^{-im\phi} \mathrm{d}\phi.
\label{gaussint}
\end{equation}

\begin{figure}[htb]
\centerline{\includegraphics[width=7cm, trim = 0in 0.8in 0in 0.6in, clip = true]{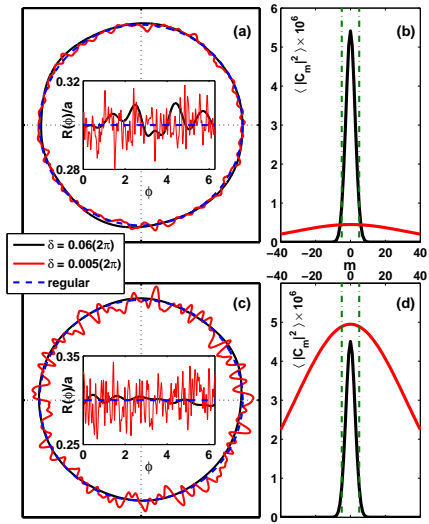}}
\caption{(Color online) (a): Polar and cartesian plots of a sample hole shape for two different correlation angles, assuming a fixed radius fluctuation $\sigma = 0.006a$; (b): The underlying distributions of $\langle |C_m|^2 \rangle$; (c), (d): same, assuming a fixed area fluctuation $\langle \delta A^2 \rangle = 0.004a^2$}
\label{holes}
\end{figure}

Within this scheme, the quantities $\sigma$, $\delta$ and $\langle C_0\rangle$ are still free parameters. To set them, we consider two different models. The first model consists in assuming, for varying $\delta$, a given magnitude of the fluctuation of the hole radius, namely by setting $\langle \delta R \rangle = 0$ and $\langle \delta R^2 \rangle = const$, which sets $\langle C_{0} \rangle = 0$ and $\sigma = \sqrt{\langle \delta R^2 \rangle}$. Fig. \ref{holes} (a) illustrates one realization of a hole within this model with $\sigma = 0.006a$, computed for two different values of $\delta$. The second model consists instead in assuming a given magnitude of the fluctuations in the hole area $A$, thus setting $\langle \delta A \rangle = 0$ and $\langle \delta A^2 \rangle = const$. The two conditions determine implicitly the parameters $\langle C_{0} \rangle$ and $\sigma$, since

\begin{equation}
\delta A = \pi\left(\sum|C_m|^2 + 2RC_0\right).
\label{deltaA}
\end{equation}

In panel (c) of Fig. \ref{holes}, we show a hole realization when those conditions are imposed, with $\langle \delta A^2 \rangle = 0.004a^2$, for the same two values of $\delta$ as in panel (a). It can be seen that in this case the magnitude of the radius fluctuations $\langle \delta R^2 \rangle$ depends substantially on $\delta$, and while the profile for $\delta = 0.06(2\pi)$ appears as a reasonable representation of what could be expected from a state-of-the-art PHC \cite{houdre, skorobogatiy}, the $\delta = 0.005(2\pi)$ profile displays unrealistically large radius fluctuations. 

\begin{figure}[htb]
\centerline{\includegraphics[width=8cm, trim = 0in 0.5in 0in 0.4in, clip = true]{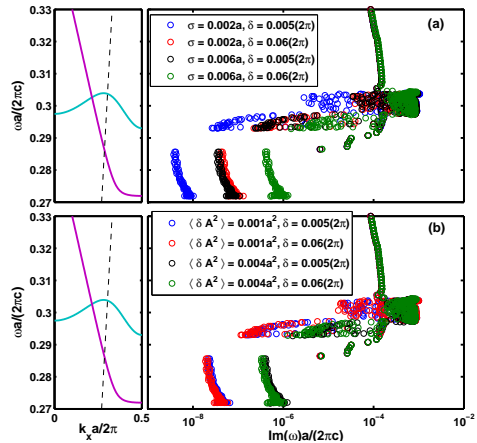}}
\caption{(Color online) Computed radiation loss rates. (a): $\langle \delta R^2 \rangle = const$ model; (b): $\langle \delta A^2 \rangle = const$ model. On the left, the guided bands of the regular structure (solid lines) and the light cone (dashed line) are displayed for reference. }
\label{rates}
\end{figure}

In Fig. \ref{rates} we plot the radiative loss rates and frequencies of the guided modes, computed for one realization of a waveguide of length $256a$. Panel (a) represents the fixed-$\langle \delta R^2 \rangle$ model, with the four possible combinations of $\delta = 0.06(2\pi)$, $\delta = 0.005(2\pi)$, $\sigma = 0.006a$ and $\sigma = 0.002a$. As can be seen, within this model, changing $\delta$ can have an effect as dramatic as that of changing $\sigma$, with an order of magnitude difference present in the loss rates of the modes when going from the high $\delta = 0.06(2\pi)$ to the low $\delta = 0.005(2\pi)$. In panel (b) we show a similar plot but for the fixed-$\langle \delta A^2 \rangle$ model, with the same two values of $\delta$ and with $\langle \delta A^2 \rangle = 0.004a^2$ and $\langle \delta A^2 \rangle = 0.001a^2$. Here, in contrast, changing $\delta$ does not have a very pronounced effect, with the radiative rates being only very slightly higher in the low-$\delta$ case. 
This single-realization data, together with the sample profiles given in Fig. \ref{holes}, suggest that fine features in the shape of the holes do not contribute as strongly to the disorder effects as do the smooth features.

\begin{figure}[htb]
\centerline{\includegraphics[width=8cm, trim = 0in 0.5in 0in 0.4in, clip = true]{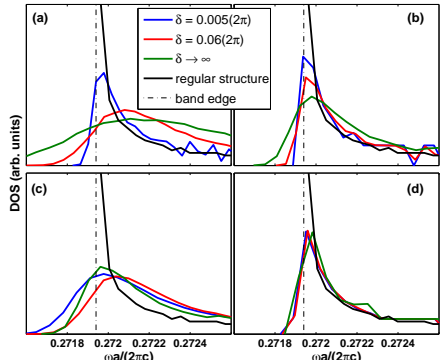}}
\caption{(Color online) DOS histograms close to the band edge (marked by a dashed-dotted line), for (a): $\sigma = 0.006a$; (b): $\sigma = 0.002a$; (c): $\langle \delta A^2 \rangle = 0.004a^2$; (d): $\langle \delta A^2 \rangle = 0.001a^2$.}
\label{DOS}
\end{figure}

To test this further, we computed the density of states (DOS) of the modes close to the band edge, by a statistical average over 500 disorder realizations. The computed DOS are plotted in Fig. \ref{DOS}, panels (a)-(b) for the fixed-$\langle \delta R^2 \rangle$ model and panels (c)-(d) for the fixed-$\langle \delta A^2 \rangle$ one. To emphasize the role of the irregular hole shape, we also plot the DOS with the commonly adopted \cite{Gerace_OL_2004, Mazoyer_PRL_2009, Savona_PRB_2011, Portalupi_PRB_2011,Noda_OE_2011} simple disorder model of constant radius fluctuations (corresponding to the limiting case $\delta\rightarrow\infty$). Oscillations in the high-frequency tails originate from the finite length of the simulated waveguide. All the computed histograms show a Lifshitz tail below the band edge. Modes in this region are spatially localized \cite{Savona_PRB_2011, arXiv}, and slow-light propagation is consequently hindered \cite{Mazoyer_PRL_2009, Mazoyer_OE_2010, Le_Thomas_PRB_2009}. From the DOS it appears, once more, that the correlation angle $\delta$ has a strong effect on the distribution only in the $\langle \delta R^2 \rangle = const$ case, where the broadening of the band edge increases considerably for increasing $\delta$.

In all our simulations, the sum in Eq. (\ref{epsprime}) was restricted to $|m| \le m_{max}$. We have carefully checked that all results were well converged starting at $m_{max} = 5$. This remark corroborates our statement that fine features in the disorder are not as relevant as the smooth ones. Intuitively, this is due to the typical wavelength of the electromagnetic modes of the PHC, which produces a spatial averaging of features in the dielectric profile on a much smaller spatial scale.

The main result of this work can be broken down in two statements. First -- confirming the suggestion made in a recent experimental study \cite{houdre} -- it is indeed the magnitude of the fluctuations of the hole area, $\langle \delta A^2 \rangle$, which determines the magnitude of the disorder effects. This is evident from Fig. \ref{rates} and Fig. \ref{DOS}, where a $\langle \delta A^2 \rangle = const$ model is virtually independent of $\delta$. Second, when $\langle \delta R^2 \rangle = const$ is imposed instead, the disorder effects are strongly dependent on $\delta$, with a small correlation angle corresponding to much smaller loss rates and spectral broadening. The effect on loss rates was already indicated by a perturbative analysis \cite{Hughes_PRL_2005}. Insight is provided by comparing Figs. \ref{holes} (b) and (d), where the average quantities $\langle |C_m|^2 \rangle$ are plotted for the corresponding hole models in panels (a) and (c). The dashed-dotted lines in panels (b) and (d) mark the convergence range $|m| \le m_{max} = 5$. In the case of fixed radius fluctuation depicted in panel (b), the values of $\langle |C_m|^2 \rangle$ in this small-$|m|$ range vary considerably when varying $\delta$. This is due to the fact that the parameters $\sigma$ and $\delta$ determine respectively the integral and the width of the distribution $\langle |C_m|^2 \rangle$ as a function of $m$. In the case of fixed hole-area fluctuations shown in panel (d) instead, the two curves differ mainly in their width, while their values in the vicinity of $m=0$ are closer than in the previous case. Thus, for fixed area fluctuations, the $C_m$ coefficients relevant to the results vary less as a function of $\delta$.

The fabrication process of a PHC does not correspond to either of the two limiting models considered in this work. The cross section of the e-beam used for lithography and the subsequent etching process are likely to set a {\em typical length scale} for the hole shape. For PHCs with nominally constant hole radii, this situation is however expected to be closer to the fixed-$\langle \delta R^2 \rangle$ assumption, as Figs. \ref{holes} (a) and (c) also clearly suggest. We conclude that, while hole-area fluctuations are the single relevant parameter to quantify disorder effects, this parameter is only indirectly determined by the details of the fabrication process, which instead directly set the amplitude of radius fluctuations and their correlation angle. If any control over these two parameters is possible in practice, then the best PHC quality will arise from the finest roughness of the irregular hole profile. This result provides a valuable indication for further improvement of the quality of PHCs in view of applications in photonics. 

%\pagebreak


\begin{thebibliography}{10}
\newcommand{\enquote}[1]{``#1''}


\bibitem{Joannopoulos_2008}
J.~D. Joannopoulos, S.~G. Johnson, J.~N. Winn, and R.~D. Meade, \emph{Photonic
  Crystals: Molding the Flow of Light} (Princeton University Press, 2008).

\bibitem{Gerace_PAN_2005}
D.~Gerace and L.~C. Andreani, Photonics Nanostruct. Fundam. Appl. \textbf{3},
  120 (2005).

\bibitem{Noda_OE_2011}
Y.~Taguchi, Y.~Takahashi, Y.~Sato, T.~Asano, and S.~Noda, Opt. Express
  \textbf{19}, 11916 (2011).

\bibitem{Portalupi_PRB_2011}
S.~L. Portalupi, M.~Galli, M.~Belotti, L.~C. Andreani, T.~F. Krauss, and
  L.~O’Faolain, Phys. Rev. B \textbf{84}, 045423 (2011).

\bibitem{Gerace_OL_2004}
D.~Gerace and L.~C. Andreani, Optics Letters \textbf{29}, 1897 (2004).

\bibitem{Hughes_PRL_2005}
S.~Hughes, L.~Ramunno, J.~F. Young, and J.~E. Sipe, Phys. Rev. Lett.
  \textbf{94}, 033903 (2005).

\bibitem{Hughes_PRB_2009}
M.~Patterson, S.~Hughes, S.~Schulz, D.~M. Beggs, T.~P. White, L.~O’Faolain,
  and T.~F. Krauss, Phys. Rev. B \textbf{80}, 195305 (2009).

\bibitem{Mazoyer_PRL_2009}
S.~Mazoyer, J.~P. Hugonin, and P.~Lalanne, Phys. Rev. Lett. \textbf{103},
  063903 (2009).

\bibitem{John_PRL_1987}
S.~John, Phys. Rev. Lett. \textbf{58}, 2486 (1987).

\bibitem{Savona_PRB_2011}
V.~Savona, Phys. Rev. B \textbf{83}, 085301 (2011).

\bibitem{Hughes_PRB_2010}
M.~Patterson and S.~Hughes, Phys. Rev. B \textbf{81}, 245321 (2010).

\bibitem{houdre}
N.~{Le Thomas}, Z.~Diao, H.~Zhang, and R.~Houdre, J. Vac. Sci. Technol. B
  \textbf{29}, 051601 (2011).

\bibitem{Mazoyer_OE_2010}
S.~Mazoyer, P.~Lalanne, J.~Rodier, J.~Hugonin, M.~Spasenović, L.~Kuipers,
  D.~Beggs, and T.~Krauss, Opt. Express \textbf{18}, 14654 (2010).

\bibitem{Le_Thomas_PRB_2009}
N.~{Le Thomas}, H.~Zhang, J.~J{\'a}gersk{\'a}, V.~Zabelin, R.~Houdre,
  I.~Sagnes, and A.~Talneau, Phys. Rev. B \textbf{80}, 125332 (2009).

\bibitem{skorobogatiy}
M.~Skorobogatiy, G.~B{\'e}gin, and A.~Talneau, Opt. Express \textbf{13}, 2487
  (2005).

\bibitem{andreani_2006}
L.~C. Andreani and D.~Gerace, Phys. Rev. B \textbf{73} (2006).

\bibitem{arXiv}
S.~R. Huisman, G.~Ctistis, S.~Stobbe, A.~P. Mosk, J.~L. Herek, A.~Lagendijk,
  P.~Lodahl, W.~L. Vos, and P.~W.~H. Pinkse, ArXiv:1201.0624v1.

\end{thebibliography}
\end{document}